\documentclass[aps,pra,preprint,su
perscriptaddress,showkeys]{revtex4-2}
\usepackage{hyperref}
\usepackage{latexsym}
\usepackage{amsfonts}
\usepackage{amssymb}
\usepackage{amsmath}
\usepackage{array}
\usepackage{dcolumn}
\usepackage{longtable}
\usepackage{booktabs}
\usepackage{graphicx}
\usepackage{color}
\usepackage[normalem]{ulem}
\usepackage{xcolor}
\usepackage{soul,xcolor}
\usepackage{float}
\usepackage{datetime}

\def\be{\begin{equation}}
\def\ee{\end{equation}}
\def\({\left(}
\def\){\right)}

\def\e{{\rm e}}

\def\be{\begin{equation}}
\def\ee{\end{equation}}

\begin{document}

\title{Free energies of triply and quadruply degenerate electron ladder spectra} 

\author{Minghao Li}
\affiliation{School of Science, Tianjin University, 92 Weijin Road, Tianjin 
300072, China}

\author{Zehua Sun}
\affiliation{Department of Physics, University of Toronto, 
60 St. George Street, Toronto, ON M5S 1A7, Canada}

\author{Klavs Hansen}
\affiliation{Center for Joint Quantum Studies and Department of Physics, 
School of Science, Tianjin University, 92 Weijin Road, Tianjin 300072, China}
\email{klavshansen@tju.edu.cn}

\date{\currenttime, ~  \today}

\begin{abstract}
The electronic free energies in metal clusters are calculated based on
a simple ladder spectrum for the level degeneracies 
three and four, and compared with the known results for degeneracies one 
and two.
The low temperature and the asymptotic high temperature results for
free energy differences are generalized to ladder spectra of any 
degeneracy.
Remarkably, free energy differences do not approach zero at high 
energies for size-independent Fermi energies.
\end{abstract}

\maketitle

\section{\label{sec:intro}Introduction}

Thermal excitation of electrons is an important contribution to 
the thermal properties of matter, both in bulk and for 
metallic clusters and nanoparticles, in spite of the usually  
relatively small amount of energy contained in these 
excitations.
The finite size of clusters often cause spectra of 
quasi-free valence electrons in small particles to have gaps
that can exceed temperatures by a significant factor.
Nevertheless, these 
excitations will still play an important part in the suppression 
of size dependent phenomena, such as electronic shell structure 
\cite{GenzkenPRL1991} or the odd-even effect \cite{Brenner1992}.
Another, related but so far unexplored effect is the thermal 
suppression of the quantized conductance in one-dimensional 
conductors, where the quantization of the transverse motion 
induces both shells and supershells \cite{YansonPRL2000} and 
odd-even effects \cite{YamaguchiSSC1997} in analogy to metal
clusters \cite{PedersenNature1991,SaundersPRB1985,ManninenZPD1994}.
Another effect for which thermally excited electronic states
is a sine qua non and which has been explored in some detail is 
the emission of light from thermally excited photo-active 
electronic states of 
gas phase molecules \cite{Andersen1996,AndersenEPJD2001,toker07,
martin13,EbaraPRL2016,FerrariIRPC2019,SaitoPRA2020}.
Other effects have been treated in the early works of 
\cite{Kubo,DentonPRB1973}.

The treatment of the thermal behavior of electrons in most of 
these situations describing free particles can be considered 
canonical even if the entire particle is not in equilibrium 
with an external heat bath that defines a macroscopic temperature.
This description is a consequence of the small electronic heat 
capacity compared to that of the vibrational motion.
This makes the vibrational degrees of freedom effectively act 
as a heat bath, even though the entire 
system is rigorously microcanonical, as already noted in 
Ref. \cite{BrackZPD1991}.
The considerations require that electronic and vibrational 
energies can be exchanged on timescales shorter than the 
experimental relevant ones, which will be assumed here.
Although there has been observations at low temperatures of 
cases where this requirement is not fulfilled, it is expected 
to be the case more often than not, and we will therefore 
consider the canonical partition functions and corresponding 
free energies in this work.

One of the original motivations for approaching the question of 
electronic free energies with the level scheme used here
was the question of the temperature dependence of the odd-even 
effect in simple metal cluster.
The question was raised in connection with the abundance pattern 
in sodium clusters, and was later addressed for gold clusters, 
which display very strong odd-even effects in several quantities.
The original expectation for the sodium clusters was that the 
effect would disappear asymptotically with temperature, when 
calculated in the framework of a doubly degenerate (equidistant 
spacing) ladder spectrum.
This, however, was shown in \cite{HansenCP2020} not to be the 
case.
In fact, half the zero temperature Fermi energy difference
between odd and even clusters was retained in the free energy 
in the high temperature limit.
It is therefore of obvious interest to know what the analogous 
values are for larger degeneracies.

Here we will use the equidistant and independent particle spectrum,
equivalent to a harmonic oscillator (or ladder) spectrum. 
The description in terms of independent particles may seem crude,
but it is worth to consider how many situations this starting point
is made in a statistical treatment, at least as an initial 
approximation. 
It is obviously required due to the huge number of states one 
would otherwise need to calculate to account for any appreciable 
entropy.
The equidistant level spectrum is clearly a schematic representation 
of real particles.
However, the Fermi gas description that works so well on a number
of bulk metallic elements, the alkalis in particular, builds 
successfully on this description.
A description of the shell structure of clusters of these metals 
also reproduces observed shell structure, from the smallest up to 
up to very large sizes.
It has likewise been found to represent gold cluster odd-even 
staggering caused by the spin degrees of freedom to a surprisingly 
good accuracy, as judged from the size-dependence of measured 
dissociation energies in these clusters \cite{HansenCP2020}.
We will therefore apply this scheme without any modifications.
Finally, in order for the level scheme to mimic the behavior of 
a Fermi gas in a particle, 
levels are lowered systematically as sizes increase in order to
keep the Fermi energy constant.

One aspect of the level structure calculated should be mentioned, 
viz. the splitting of levels due to Jahn-Teller deformations.
These are not here.
To do so would make the theory system specific, yet without being 
entirely realistic, and it would tend to hide the features the 
calculations are trying to shed light on, which are the 
asymptotically high temperature free energy differences  
mentioned above for the degeneracy two, $g=2$, case.
The role of deformations will be considered in the discussion 
section.

\section{Computational procedure}

The calculation of canonical partition functions for any  
highly degenerate fermionic systems is complicated due to 
the combined effect of particle number conservation and 
Pauli blocking.
Several methods have been devised to perform these and related
calculations \cite{WilliamsNPA1969,BrackZPD1991,BorrmannJCP1993}.
One method is completely numerical and is based on a two 
dimensional recurrence where the recurrence variables are 
the electron numbers and single particle state energies.
The equation reads \cite{BrackZPD1991}
\begin{equation}
\label{iterative}
Z(n_l,N)=Z(n_l-1,N)+e^{-\beta\epsilon_{n_l}}Z(n_l-1,N-1),
\end{equation}
where $n_l$ numbers the state, and the energy is energy $\epsilon_{n_l}$
(we use units where $k_{\rm B}$ is equal to one).
A similar equation was suggested for calculating nuclear level densities in 
\cite{WilliamsNPA1969}.
The method in Eq.(\ref{iterative}) will be used here as a check 
of the analytical results obtained.

\begin{figure}[htbp]
\includegraphics[width=10cm]{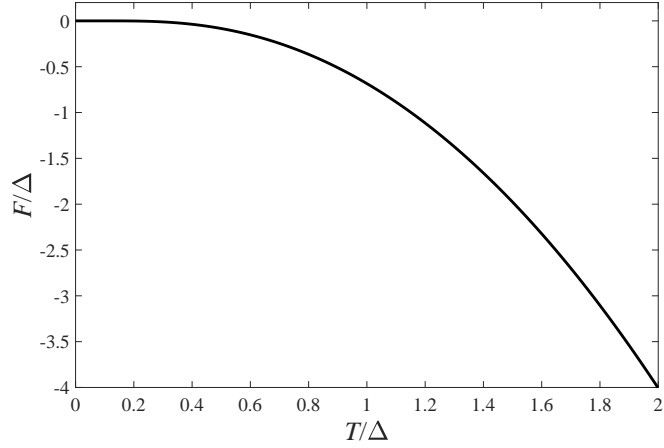}
\caption{\label{one} The free energy of the singly degenerate 
($g=1$) ladder.}
\end{figure}

Another method, specific to the equidistant spectra here, 
proceeds by convoluting $g$ partition functions for a ladder of
degeneracy one to get the total degeneracy $g$ spectrum.
The calculation requires an expression for the partition function
for the singly degenerate ladder spectrum.
The ladder spectrum has energies of the single particle states of
\be
E_n = \Delta n.
\ee
This is known, and is easily calculated with a recurrence relation
based on the separation of the sum over configurations into those
that have the lowest single particle state occupied and those where
it is unoccupied, see e.g. \cite{SchonhammerAJP2000}.
The result for $N$ electrons is
\be
Z_1 = \prod_{k=1}^N \(1-\e^{-\beta k \Delta}\)^{-1},
\ee
where the total ground state energy has been set to zero.
The free energy of this is shown in Fig. \ref{one}.

For the degeneracy $g=2$ cases there are two different situations 
to consider, corresponding to an odd and an even electron number. 
Both are calculated in the limit of $N \rightarrow \infty$ by adding 
all possible ground state configurations weighted by the proper
Boltzmann factor, and multiplied by $Z_1^2$. 
The results are \cite{HansenCP2020}, with both ground state energies 
set to zero,
\begin{eqnarray}
Z_2^{\rm e} &=& Z_1^2 \sum_{m=-\infty}^{\infty} \e^{-\beta \Delta m^2} \\
Z_2^{\rm o} &=& Z_1^2 \sum_{m=-\infty}^{\infty} \e^{-\beta \Delta m(m-1)}.
\end{eqnarray}

\begin{figure}[H]
\centering
\includegraphics[width=10cm]{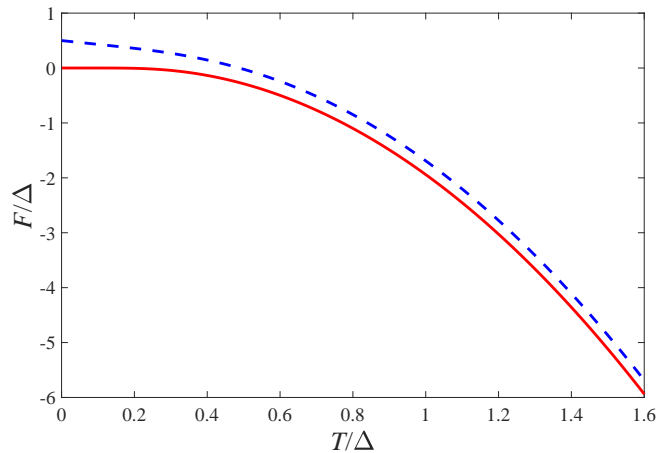}
\caption{\label{two} 
The free energies of the doubly degenerate ladder for odd (dashed 
line) and even (full line) electron numbers.}
\end{figure}

\begin{figure}[H]
\centering
\includegraphics[width=10cm]{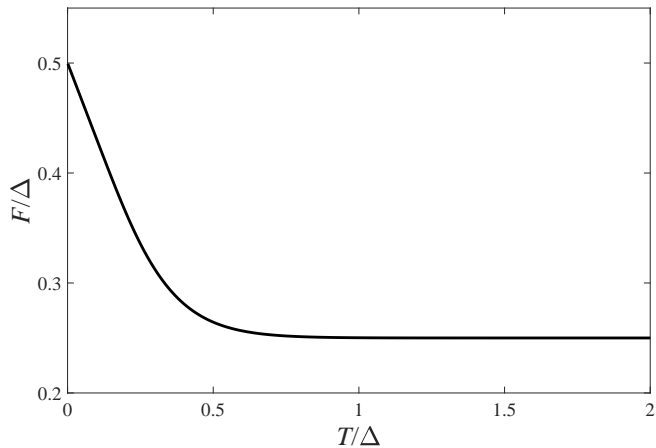}
\caption{\label{two-2} The difference in free energies ($F_2^{\rm o}-F_2^{\rm e}$)
between the odd and the even electron number degenerate ladders.}
\end{figure}

However, adding one electron (arriving with a monovalent atom) to a 
particle will lower the levels.
We assume here and for the other cases that this lowering occurs
linearly with the number of electrons added.  
Keeping the energy of the highest occupied orbitals of systems 
$N$ and $N+2$ identical determines the shift in energy and the 
additional, relative Boltzmann factor 
between the even and the odd electron number particle.
We have: 
\begin{eqnarray}
Z_2^{\rm e} &=& Z_1^2 \sum_{m=-\infty}^{\infty} \e^{-\beta \Delta m^2}\\
Z_2^{\rm o} &=& Z_1^2 \e^{-\beta \Delta/2} 
\sum_{m=-\infty}^{\infty} \e^{-\beta \Delta m(m-1)}.
\end{eqnarray}
The corresponding free energies are shown in Fig. \ref{two}, and 
Fig. \ref{two-2} shows the difference of the free energies.

One particularly interesting result is the fact that the free 
energy difference does not approach zero for high temperatures, 
but converges to $\Delta/4$, i.e. half the zero temperature 
difference.
Another interesting result is that this asymptotic value is 
approached at temperatures significantly below $\Delta$. 

For a general value of the degeneracy the expression for the 
partition function is
\be
Z = Z_1^g  \sum_{(n_1,n_2,...n_g)} \e^{-\beta E^{(0)}(n_1,n_2,...,n_g)},
\ee
where $n_i$ is now the number of electrons in ladder number $i$.
The summation ranges over all possible values of the set of $n$'s,
i.e. non-negative values that obey the constraint 
\be
\sum_i n_i = N.
\ee
The total ground state energy for any specific distribution of 
electrons on the $g$ different ladders is then, with the zero of 
energy chosen as the common ground state energy for the single 
particle states, equal to
\be
E^{(0)}(n_1,n_2,...,n_g) = \Delta \sum_{i=1}^g \frac{1}{2}n_i (n_i-1).
\ee 
For $g=3$ this is
\be
E^{(0)}(n_1,n_2,n_3) = \frac{\Delta}{2} 
\left[ n_1 (n_1-1)+n_2 (n_2-1) + n_3 (n_3-1)\right].
\ee
The fixed particle number is implemented by eliminating $n_3$ 
with the relation
\be
N = n_1 + n_2  + n_3,
\ee
to get
\be
E^{(0)}(n_1,n_2) = \frac{\Delta}{2}\left[2n_1^2 + 
2 n_2^2 +N^2-N +2n_1n_2-2N(n_1+n_2)  \right]. 
\ee
In this expression $n_1$ and $n_2$ can take any non-negative 
values as long as their sum does not exceed $N$,
\be
n_1 + n_2 \leq N.
\ee
The energy has minimum around $n_1,n_2 = (N/3,N/3)$, possibly 
with a small degeneracy, and hence we also have the minimum 
for $n_3$ around $N/3$.

The value of $N$ is arbitrary, in the sense that as long as it is 
large enough it will not matter precisely how large. 
It therefore makes sense to eliminate this number.
This is done by redefining the occupation numbers by subtraction 
of the values pertaining to the absolute ground state.
We consider the values $N=3M+f$, with $M$ a large integer, where 
$f$ takes the values 0, 1 or 2, corresponding to a filled shell, or 
1 or 2 additional electrons.
Irrespective of the value of $f$ we subtract $M$ from all $n_i$.
This gives
\be
E^{(0)}(n_1,n_2) = \frac{\Delta}{2}
\left[2n_1^2+2n_2^2+2n_1n_2-2f(n_1+n_2) -f+f^2 +3M(M-1) +2Mf \right]. 
\ee
From this we subtract the lowest energy which is attained 
at $n_1=n_2=0$ for $f=0$, for $f=1$ at the three sets 
$(n_1,n_2)=(0,0),(1,0),(0,1)$, and for $f=2$ at the three 
values $(n_1,n_2)=(1,0),(0,1),(1,1)$.
These energies are zero if the term involving $M$ are 
subtracted, which we will then do.
We then have with the redefined occupation numbers and 
zeroes of energy the expression
\be
E^{(0)}(n_1,n_2) = \frac{\Delta}{2}
\left[2n_1^2+2n_2^2+2n_1n_2-2f(n_1+n_2) -f+f^2 \right]. 
\ee

\begin{figure}[H]
	\centering
	\includegraphics[width=10cm]{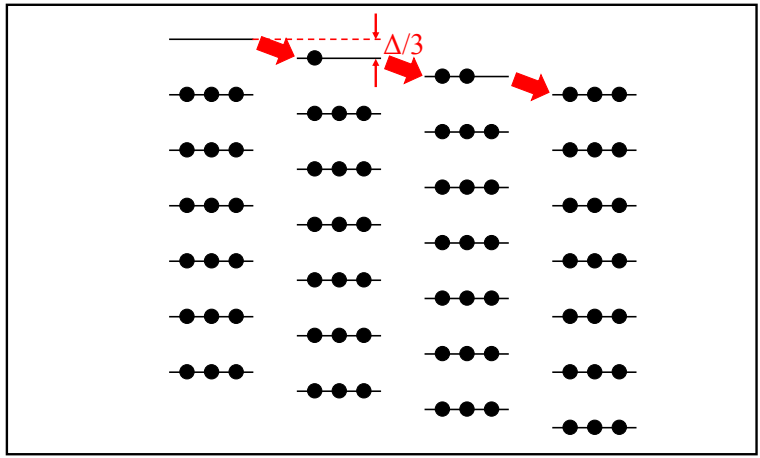}
	\caption{\label{lowering3} The systemats of the 
	lowering of levels for the triply degenerated ladders. 
	Dots indicate the population of one of the ground states. 
	All lines can accommodate three electrons.}
\end{figure}

To continue, we complete the squares in this expression.
Chose $n_1$ as the first. 
After completing the square also for $n_2$, we have
\be
E^{(0)}(n_1,n_2) = \Delta \( \(n_1+\frac{1}{2}\(n_2-f \) \)^2 
+ \frac{3}{4}\(n_2-\frac{1}{3}f\)^2
+ \frac{f^2}{6}-\frac{f}{2}\).
\ee
As a check we see that the lowest energies are 0 for all 
three values of $f$.
As mentioned above, the ground state energy should change 
with $f$. 
The variation corresponds to an additional contribution of
$\Delta(f - f^2/3)$.
Fig. \ref{lowering3} shows the systematically lowered levels 
for the triply degenerated ladders.
The resulting energy is then
\be
E^{(0)}(n_1,n_2) = \Delta \( \(n_1+\frac{1}{2}\(n_2-f \)\)^2 
+ \frac{3}{4}\(n_2-\frac{1}{3}f\)^2
+\frac{1}{2}f-\frac{1}{6}f^2\).
\ee

We can now write down the partition function in terms of summations 
over $n_1$ and $n_2$:
\be
Z_3 = Z_1^3  \e^{-\beta \Delta\(f/2-f^2/6\)}
\sum_{n_1,n_2=-\infty}^{\infty} 
\exp\( -\beta \Delta  \left[ \(n_1+\frac{1}{2}\(n_2-f \) \)^2 
+ \frac{3}{4}\(n_2-\frac{1}{3}f\)^2 \right] \)
\ee
The calculation of the sums in this expression depends on the 
value of $f$, and the summation over $n_1$ also depends on $n_2$. 
This gives two contributing sum for each value of $f$.
One by one they are:

$f=0, n_2$ odd ($n_2=2n_2'+1$):\\
The term $(n_2-f)/2$ in the first bracket is then an integer 
plus 1/2.
As we sum over all integer values of $n_1$, the integer part 
of this term can be set to zero. 
Eliminating the apostrophe in the following, the contribution 
from this sector to the sum is then  
\begin{eqnarray}
\sum_{n_1,n_2=-\infty}^{\infty} 
\exp\( -\beta \Delta  
\left[ \(n_1+n_2+\frac{1}{2}\)^2 + 3\(n_2+\frac{1}{2} \)^2 \right] \) ,
\end{eqnarray}
where $n_2$ runs over all integers here and in the following. 
Thus:
\begin{eqnarray}
	\sum_{n_1=-\infty}^{\infty} 
	\exp\left[ -\beta \Delta \(n_1+\frac{1}{2}\)^2 \right]
	\sum_{n_2=-\infty}^{\infty} 
	\exp\left[ -3\beta\Delta\(n_2+\frac{1}{2}\)^2\right],
\end{eqnarray}

The contribution from $f=0, n_2$ even is calculated similarly to 
give
\begin{eqnarray}
\sum_{n_1=-\infty}^{\infty} 
\exp\( -\beta \Delta n_1^2 \)
\sum_{n_2=-\infty}^{\infty} 
\exp\( -3\beta\Delta n_2^2 \),
\end{eqnarray}

To simplify the notation we will define a shorthand for the sums 
as
\be
\sigma(a,b) \equiv \sum_{n = -\infty}^{\infty} 
\exp\(- a \beta \Delta (n+b)^2 \),
\ee
where the dependence on $\beta \Delta$ is implicit.
This gives
\begin{eqnarray}
Z_3 (f=0)&=& Z_1^3 \left[\sigma(1,0)\sigma(3,0)
+\sigma\(1,\frac{1}{2}\)\sigma\(3,\frac{1}{2}\)\right].
\end{eqnarray}
For $f=1$ we have
\begin{eqnarray}
Z_3 (f=1)&=& \e^{-\beta \Delta /3} Z_1^3 \left[
\sigma(1,0)\sigma\(3,\frac{1}{3}\) + \sigma\(1,\frac{1}{2}\)\sigma\(3,-\frac{1}{6}\)
\right],
\end{eqnarray}
and for $f=2$ the sums become
\begin{eqnarray}
Z_3 (f=2)&=& \e^{-\beta \Delta /3} Z_1^3 \left[
\sigma\(1,\frac{1}{2}\)\sigma\(3,\frac{1}{6}\)
+\sigma\(1,0\)\sigma\(3,-\frac{1}{3}\)\right].
\end{eqnarray}

\begin{figure}[H]
	\centering
	\includegraphics[width=10cm]{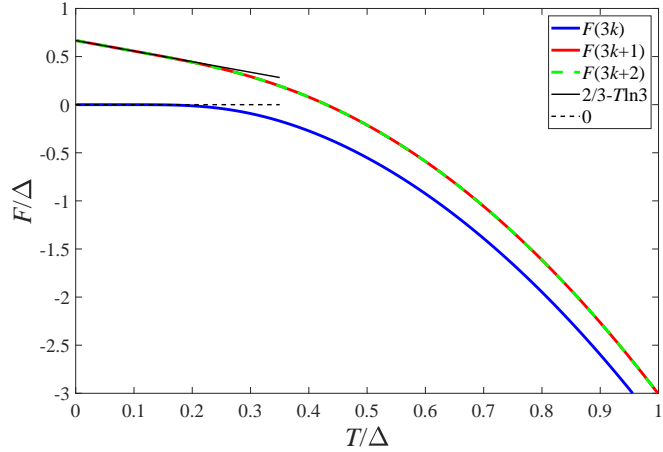}
	\caption{\label{three}
	The free energies of the triply degenerate ladder.
	The lowest (blue) curve gives the $f=0$ case, and the
	upper are the two numerical identical curves for $f=1$ and $f=2$.
	The short straight lines are the first order temperature approximations.}
\end{figure}

\begin{figure}[H]
	\centering
	\includegraphics[width=10cm]{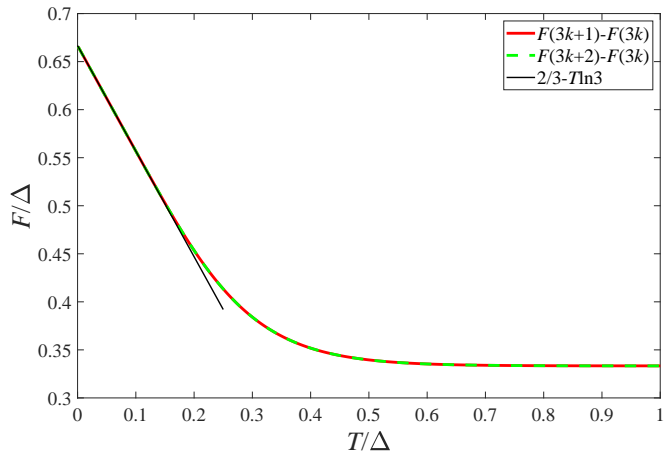}
	\caption{\label{three-2} The free energy difference between the 
	$f=1,2$ ladders and the closed shell, $f=0$, ladder systems.
The straight line is the first order temperature approximation of the
difference.}
\end{figure}

Fig. \ref{three} shows the free energies calculated with these
equations, performing the summation numerically. 
Fig. \ref{three-2} shows the difference of the free energies of the 
partially occupied systems relative to the closed shell, $f=0$, system.
Figs. \ref{three},\ref{three-2} also show the free energy and free 
energy difference calculated to first order in the temperature.
The approximations, $F = E(0)-TS(0)$, are calculated with the 
entropies calculated from the ground state degeneracies which are 
given by the combinatorial factors $C_3^k, k=0,1,2$:
\begin{eqnarray}
	S(T=0,f=0) &=& \ln(C_3^0)=0 \\
	S(T=0,f=1) &=& \ln(C_3^1)=\ln3 \\
	S(T=0,f=2) &=& \ln(C_3^2)=\ln3.		
\end{eqnarray}

\begin{figure}[H]
	\centering
	\includegraphics[width=10cm]{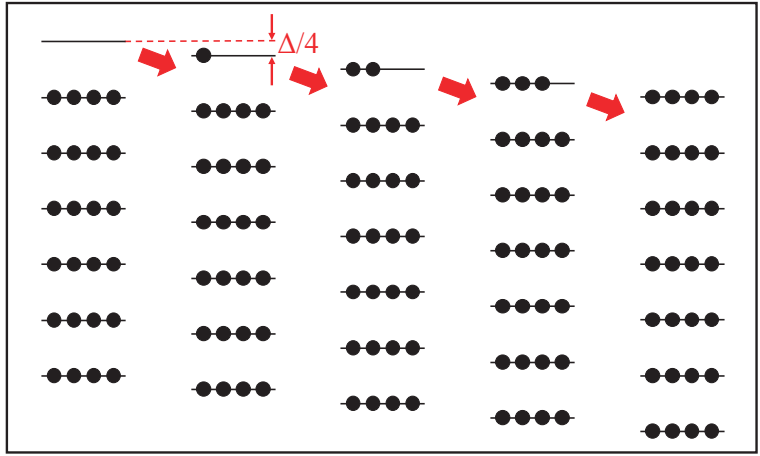}
	\caption{\label{lowering4} The lowering of levels for 
	the quadruply degenerated ladders, similar to the 
	the $g=3$ case with the difference that lines can 
	accommodate four electrons here.}
\end{figure}

For $g=4$ the ground state energies calculated analogously take the
form
\begin{eqnarray}
E\(n_1,n_2,n_3\) &=& \frac{\Delta}{2} \{
2\(n_1^2+n_2^2+n_3^2\)
+2\(n_1n_2+n_2n_3+n_3n_1\) \\\nonumber
&-&
2f\(n_1+n_2+n_3\)
+4M\(M-1\)
+2Mf
+f\(f-1\)\},
\end{eqnarray}
with the total number of electrons equal to $4M+f$.
Adding $f(f-1)/2$ to this makes the energy zero for all values of $f$:
\begin{eqnarray}
	E\(n_1,n_2,n_3\)	
	=
	E^{(0)} &+& \frac{f\(f-1\)}{2}\Delta \\\nonumber	
	=
	E^{(0)} &+& \Delta
	\biggl[
	\(n_1+\frac{n_2}{2}+\frac{n_3}{2}-\frac{f}{2}\)^2 +
	\frac{3}{4}\(n_2+\frac{n_3}{2}-\frac{f}{3}\)^2 \\\nonumber
	&+&
	\frac{2}{3}\(n_3-\frac{f}{4}\)^2 +
	\frac{f^2}{8}-\frac{f}{2} \biggr].
\end{eqnarray}
Similar to the procedure used for the $g=3$ case, the ground state 
energy varies with $f$.
This gives an additional contribution of $f-f^2/4$ which, notably, 
is also minus twice the term already present in $f$, similar to the 
$g=3$ case.
Fig.\ref{lowering4} shows the systematically lowered levels for the 
quadruply degenerated ladders.
The resulting energy is then:
\begin{eqnarray}
\label{E4}
	E\(n_1,n_2,n_3\)	
	&=&
	\Delta
	\biggl[
	\(n_1+\frac{n_2}{2}+\frac{n_3}{2}-\frac{f}{2}\)^2 +
	\frac{3}{4}\(n_2+\frac{n_3}{2}-\frac{f}{3}\)^2 \\\nonumber
	&+&
	\frac{2}{3}\(n_3-\frac{f}{4}\)^2 -\frac{f^2}{8}+\frac{f}{2} \biggr].	
\end{eqnarray}

\begin{figure}[H]
	\centering
	\includegraphics[width=10cm]{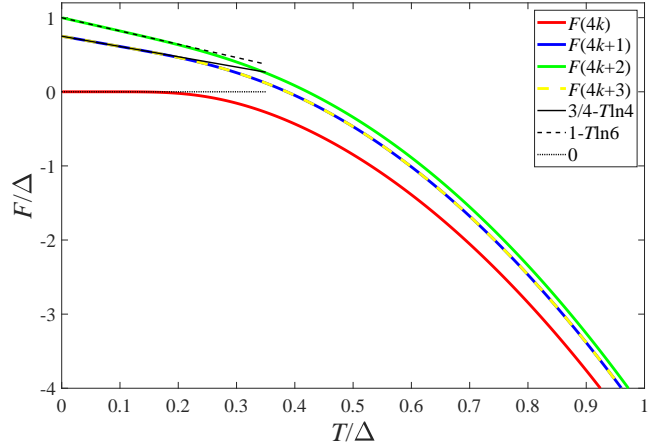}
	\caption{\label{four} The free energies of the quadruply 
	degenerate ladder.
The lowest curve is for the filled shell, $f=0$.
The middle curve is for the numerical identical curves for $f=1$
and $f=3$, and the top curve for $f=2$.}
\end{figure}

\begin{figure}[H]
	\centering
	\includegraphics[width=10cm]{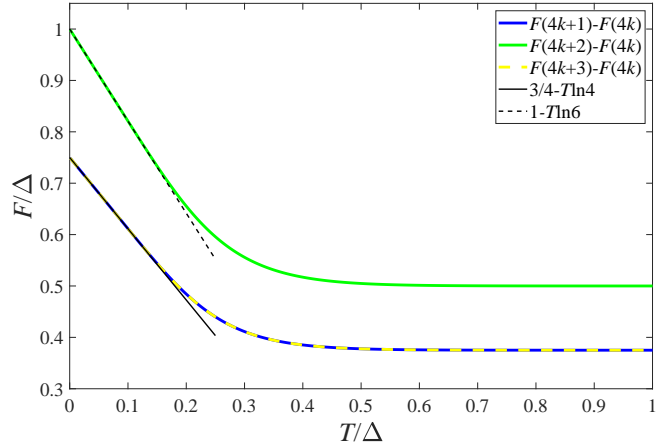}
	\caption{\label{four-2} The free energy difference between 
	the $f=1,3$ ladders (lowest curve), $f=2$ (top curve), and 
	the closed shell ladder ($f=0$).}
\end{figure}

We can now write down the partition function in terms of summations 
over $n_1$, $n_2$ and $n_3$:
\begin{eqnarray}
	Z_4=Z_1^4 e^{-\beta\Delta (f/2-f^2/8)} \sum_{n_1,n_2,n_3=-\infty}^{\infty}
	\exp\biggl\{-\beta\Delta \biggl[
	\(n_1+\frac{1}{2}n_2+\frac{1}{2}n_3-\frac{1}{2}f\)^2 \\\nonumber +
	\frac{3}{4}\(n_2+\frac{1}{3}n_3-\frac{1}{3}f\)^2 +
	\frac{2}{3}\(n_3-\frac{1}{4}\)^2
	\biggr] \biggr\}
\end{eqnarray}
For each value of $f$ this gives rise to 12 terms.
The terms for $f=1,2,3$ are listed in the SI.
The sums for $f=0$ are, with $n_1$ and the primed summation variables 
running over all integer values:\\
\begin{eqnarray}
n_2=2n_2'+1,n_3=6n_3'-2~~~~~~&&\sigma(1,\frac{1}{2})\sigma(3,\frac{1}{6})\sigma(24,-\frac{1}{3})\\
n_2=2n_2'+1,n_3=6n_3'-1~~~~~~&&\sigma(1,0)\sigma(3,\frac{1}{3})\sigma(24,-\frac{1}{6})\\
n_2=2n_2'+1,n_3=6n_3'~~~~~~&&\sigma(1,\frac{1}{2})\sigma(3,\frac{1}{2})\sigma(24,0)\\
n_2=2n_2'+1,n_3=6n_3'+1~~~~~~&&\sigma(1,0)\sigma(3,\frac{2}{3})\sigma(24,\frac{1}{6})\\
n_2=2n_2'+1,n_3=6n_3'+2~~~~~~&&\sigma(1,\frac{1}{2})\sigma(3,\frac{5}{6})\sigma(24,\frac{1}{3})\\
n_2=2n_2'+1,n_3=6n_3'+3~~~~~~&&\sigma(1,0)\sigma(3,0)\sigma(24,\frac{1}{2})\\
n_2=2n_2',n_3=6n_3'-2~~~~~~&&\sigma(1,0)\sigma(3,-\frac{1}{3})\sigma(24,-\frac{1}{3})\\
n_2=2n_2',n_3=6n_3'-1~~~~~~&&\sigma(1,\frac{1}{2})\sigma(3,-\frac{1}{6})\sigma(24,-\frac{1}{6})\\
n_2=2n_2',n_3=6n_3'~~~~~~&&\sigma(1,0)\sigma(3,0)\sigma(24,0)\\
n_2=2n_2',n_3=6n_3'+1~~~~~~&&\sigma(1,\frac{1}{2})\sigma(3,\frac{1}{6})\sigma(24,\frac{1}{6})\\
n_2=2n_2',n_3=6n_3'+2~~~~~~&&\sigma(1,0)\sigma(3,\frac{1}{3})\sigma(24,\frac{1}{3})\\
n_2=2n_2',n_3=6n_3'+3~~~~~~&&\sigma(1,\frac{1}{2})\sigma(3,\frac{1}{2})\sigma(24,\frac{1}{2})
\end{eqnarray}
These, and the analogous expressions for the three other $f$ values, 
can be condensed somewhat by noticing some relations between the 
$\sigma$ functions:
\begin{eqnarray}
\sigma(a,b) = \sigma(a,-b) = \sigma(a,1-b)\\
\sigma(a,b) + \sigma(a,b+1/2) = \sigma(a/4,2b).
\end{eqnarray}
The total partition functions for the four cases then are:
\begin{eqnarray}
	Z_4(f=0)&=& Z_1^4 (
	2\sigma(1,1/2)\sigma(3,1/6)\sigma(6,1/3)\\\nonumber 
	&+&\sigma(1,1/2)\sigma(3,1/2)\sigma(6,0)\\\nonumber
	&+&2\sigma(1,0)\sigma(3,1/3)\sigma(6,1/3)\\\nonumber
	&+&\sigma(1,0)\sigma(3,0)\sigma(6,0)),\\
	Z_4(f=1) &=& Z_1^4 e^{-\frac{3}{8}\beta\Delta}(
	\sigma(1,1/2)\sigma(3,1/6)\sigma(3/2,1/6)\\\nonumber
	&+&\sigma(1,1/2)\sigma(3,1/2)\sigma(6,1/4)\\\nonumber
	&+&\sigma(1,1/2)\sigma(3,1/3)\sigma(6,5/12)\\\nonumber
	&+&\sigma(1,0)\sigma(3,0)\sigma(6,1/4)\\\nonumber
	&+&\sigma(1,0)\sigma(3,1/3)\sigma(6,1/12)),\\
	Z_4(f=2) &=& Z_1^4 e^{-\frac{1}{2}\beta\Delta}(
	     2\sigma(1,1/2)\sigma(3,1/6)\sigma(6,1/6)\\\nonumber
	&+&2\sigma(1,1/2)\sigma(3,1/2)\sigma(24,1/4)\\\nonumber
       &+&2\sigma(1,0)\sigma(3,0)\sigma(24,1/4)\\\nonumber
     	 &+&2\sigma(1,0)\sigma(3,1/3)\sigma(6,1/6)\\
		Z_4(f=3) &=& Z_1^4 e^{-\frac{3}{8}\beta\Delta}(
  \sigma(1,1/2)\sigma(3,1/6)\sigma(3/2,1/6)\\\nonumber
	&+&\sigma(1,0)\sigma(3,1/3)\sigma(3/2,1/6)\\\nonumber
	&+&\sigma(1,0)\sigma(3,0)\sigma(6,1/4)\\\nonumber
	&+&\sigma(1,1/2)\sigma(3,1/2)\sigma(6,1/4)).
\end{eqnarray}
These expression can be used with the high temperature expansion of
the sums given in Ref. \cite{HansenCP2020}.
Here the sums will be performed numerically. 
Fig. \ref{four} shows the free energies calculated with these
equations.  
Fig. \ref{four-2} shows the difference of the free energies 
of the partially occupied systems relative to the closed shell, 
$f=0$, system.
Note that the fact that the free energy differences approach a
constant value for all is readily seen from the fact that the sum
of the reciprocal square roots of the products of the three $a$'s
is the same for all four $f$'s. 
This follows from the asymptotic values of the $\sigma$'s
which are $\sigma(a,b) \rightarrow \sqrt{\pi/a \beta \Delta}, 
T \rightarrow \infty$.

The ground state degeneracies giving rise to the low temperature 
entropy are calculated as for the $g=3$ case;
\begin{eqnarray}
	S(T=0,f=0) &=& k_B\ln(C_4^0)=0 \\
	S(T=0,f=1) &=& k_B\ln(C_4^1)=\ln4 \\
	S(T=0,f=2) &=& k_B\ln(C_4^2)=\ln6 \\
	S(T=0,f=3) &=& k_B\ln(C_4^3)=\ln4.
\end{eqnarray}
These can of course equally well be calculated by expanding 
the sums and collecting the terms with the numerically smallest
coefficients of $1/T$ in the exponential.

\section{Discussion}

Two values of the calculated free energies differences between
different values of the shell filling parameter $f$ are of 
special interest, those at zero temperature and at asymptotically 
high temperatures.
The intermediate between these two limits is to a decent 
approximation represented by a linear interpolation with a slope
given by the zero temperature entropy.
The zero temperature difference is in all cases twice the difference 
in the exponential prefactor to the sums. 
This is also clear from the derivation with the addition of the 
quadratic form in $f\Delta$, at least for the cases derived here.
We expect this will hold also for larger $g$'s, because both the 
ground state energies and the variation of the added terms that 
adjust the total energy depend only on $f$ to second order.
Basically, a linearly decreasing ladder combined with a linearly
increasing occupation number for the highest level will produce
a parabolic dependence, irrespective of the value of $g$.

The high temperature limit of the free energy differences is 
established by approximating the sums in the $\sigma$'s with the 
corresponding integrals.
It is clear from the form of Eq. (\ref{E4}) and the generalization 
to larger $g$'s that the asymptotically leading order terms will 
be identical for the sums of all values of $f$, and that this 
part of the partition function will therefore be asymptotically 
identical for all $f$.
The difference in free energies for different values of $f$ for 
same ladder spectrum will consequently only depend on the Boltzmann
prefactor involving $f$ and $\Delta$. 
As this contribution to the free energy is temperature independent,
we can simply set the value to an $f$ dependent constant plus an 
$f$-independent contribution:
\be
F \rightarrow F_{g,f} + F_g(T),~~~~  T \rightarrow \infty,
\ee
where 
\be
F_{g,f} = \frac{1}{2}\Delta \(1-f/g\)f
\ee
and
\be
F_g(T) = -\frac{g T}{2} \ln\( \frac{\pi T}{\Delta}\).
\ee
The zero temperature approximations are given by
\be
F(0) = \Delta \(1-f/g\)f.
\ee
The zero temperature is given as the logarithm of the 
relevant binomial coefficient
\be
S(0) = \ln(C_g^f).
\ee

At this point it is appropriate to discuss the effects of the 
level splitting induced by the Jahn-Teller deformations.
The effect is fundamentally a matter of energy optimization.
With finite temperatures one must instead consider free energy 
optimization.
Even in the best case where ground state deformations had been 
determined, this makes level splittings temperature dependent 
and the issue of determining the optimal geometry at finite 
temperatures one of selfconsistency. 
Adding material constants as in particular surface tension or bond 
stiffness will compound the problems.

\section{Summary}

We have calculated the Helmholtz free energy for a schematic 
metal particle with degeneracies three and four. 
The particle has a average Fermi energy, or HOMO level, which 
is independent of size, consistent with a constant density Fermi 
gas, and the results refer to this zero of energy. 
We find that the zero temperature, the low temperature and the 
high temperature differences in free energies can be described 
by simple relations determined by the gap parameter and the 
degeneracy of the levels, in addition to the number of electrons 
in the upper level.
The results generalize to a zero free energy relative to the 
filled shell ($f=0$) value of $F_g(f) - F_g(0) = \Delta/g 
\times (g-f)f$.
This values decreases with temperature approximately with the 
ground state entropy which is given by the binomial coefficient
$C_g^f$ and reaches an asymptotic value of half the $T=0$ 
difference, $\Delta/2g \times (g-f)f$. 

\section{Author contribution statements}
KH conceived the idea and calculated the $g=3$ analytical case. 
ML and ZS calculated jointly the $g=4$ case analytically and the $g=3$
and $g=4$ cases numerically, and plotted the data. 
KH wrote the paper.
%


\newpage
\begin{thebibliography}{21}%
\makeatletter
\providecommand \@ifxundefined [1]{%
 \@ifx{#1\undefined}
}%
\providecommand \@ifnum [1]{%
 \ifnum #1\expandafter \@firstoftwo
 \else \expandafter \@secondoftwo
 \fi
}%
\providecommand \@ifx [1]{%
 \ifx #1\expandafter \@firstoftwo
 \else \expandafter \@secondoftwo
 \fi
}%
\providecommand \natexlab [1]{#1}%
\providecommand \enquote  [1]{``#1''}%
\providecommand \bibnamefont  [1]{#1}%
\providecommand \bibfnamefont [1]{#1}%
\providecommand \citenamefont [1]{#1}%
\providecommand \href@noop [0]{\@secondoftwo}%
\providecommand \href [0]{\begingroup \@sanitize@url \@href}%
\providecommand \@href[1]{\@@startlink{#1}\@@href}%
\providecommand \@@href[1]{\endgroup#1\@@endlink}%
\providecommand \@sanitize@url [0]{\catcode `\\12\catcode `\$12\catcode
  `\&12\catcode `\#12\catcode `\^12\catcode `\_12\catcode `\%12\relax}%
\providecommand \@@startlink[1]{}%
\providecommand \@@endlink[0]{}%
\providecommand \url  [0]{\begingroup\@sanitize@url \@url }%
\providecommand \@url [1]{\endgroup\@href {#1}{\urlprefix }}%
\providecommand \urlprefix  [0]{URL }%
\providecommand \Eprint [0]{\href }%
\providecommand \doibase [0]{https://doi.org/}%
\providecommand \selectlanguage [0]{\@gobble}%
\providecommand \bibinfo  [0]{\@secondoftwo}%
\providecommand \bibfield  [0]{\@secondoftwo}%
\providecommand \translation [1]{[#1]}%
\providecommand \BibitemOpen [0]{}%
\providecommand \bibitemStop [0]{}%
\providecommand \bibitemNoStop [0]{.\EOS\space}%
\providecommand \EOS [0]{\spacefactor3000\relax}%
\providecommand \BibitemShut  [1]{\csname bibitem#1\endcsname}%
\let\auto@bib@innerbib\@empty
\bibitem [{\citenamefont {Genzken}\ and\ \citenamefont
  {Brack}(1991)}]{GenzkenPRL1991}%
  \BibitemOpen
  \bibfield  {author} {\bibinfo {author} {\bibfnamefont {O.}~\bibnamefont
  {Genzken}}\ and\ \bibinfo {author} {\bibfnamefont {M.}~\bibnamefont
  {Brack}},\ }\href@noop {} {\bibfield  {journal} {\bibinfo  {journal} {Phys.
  Rev. Lett.}\ }\textbf {\bibinfo {volume} {67}},\ \bibinfo {pages} {3286}
  (\bibinfo {year} {1991})}\BibitemShut {NoStop}%
\bibitem [{\citenamefont {Mottelson}(1992)}]{Brenner1992}%
  \BibitemOpen
  \bibfield  {author} {\bibinfo {author} {\bibfnamefont {B.~R.}\ \bibnamefont
  {Mottelson}},\ }in\ \href@noop {} {\emph {\bibinfo {booktitle} {Clustering
  Phenomena in Atoms and Nuclei}}},\ \bibinfo {editor} {edited by\ \bibinfo
  {editor} {\bibfnamefont {M.}~\bibnamefont {Brenner}}, \bibinfo {editor}
  {\bibnamefont {L{\"o}nnroth}},\ and\ \bibinfo {editor} {\bibfnamefont
  {F.~B.}\ \bibnamefont {Malik}}}\ (\bibinfo  {publisher} {Springer-Verlag},\
  \bibinfo {address} {Berlin},\ \bibinfo {year} {1992})\ pp.\ \bibinfo {pages}
  {571 -- 581}\BibitemShut {NoStop}%
\bibitem [{\citenamefont {Yanson}\ \emph {et~al.}(2000)\citenamefont {Yanson},
  \citenamefont {Yanson},\ and\ \citenamefont {van
  Ruitenbeek}}]{YansonPRL2000}%
  \BibitemOpen
  \bibfield  {author} {\bibinfo {author} {\bibfnamefont {A.}~\bibnamefont
  {Yanson}}, \bibinfo {author} {\bibfnamefont {I.}~\bibnamefont {Yanson}},\
  and\ \bibinfo {author} {\bibfnamefont {J.}~\bibnamefont {van Ruitenbeek}},\
  }\href {https://doi.org/10.1103/PhysRevLett.84.5832} {\bibfield  {journal}
  {\bibinfo  {journal} {Physical review letters}\ }\textbf {\bibinfo {volume}
  {84}},\ \bibinfo {pages} {5832} (\bibinfo {year} {2000})}\BibitemShut
  {NoStop}%
\bibitem [{\citenamefont {Yamaguchi}\ \emph {et~al.}(1997)\citenamefont
  {Yamaguchi}, \citenamefont {Yamada},\ and\ \citenamefont
  {Yamamoto}}]{YamaguchiSSC1997}%
  \BibitemOpen
  \bibfield  {author} {\bibinfo {author} {\bibfnamefont {F.}~\bibnamefont
  {Yamaguchi}}, \bibinfo {author} {\bibfnamefont {T.}~\bibnamefont {Yamada}},\
  and\ \bibinfo {author} {\bibfnamefont {Y.}~\bibnamefont {Yamamoto}},\
  }\href@noop {} {\bibfield  {journal} {\bibinfo  {journal} {Solid State Com.}\
  }\textbf {\bibinfo {volume} {102}},\ \bibinfo {pages} {779} (\bibinfo {year}
  {1997})}\BibitemShut {NoStop}%
\bibitem [{\citenamefont {Pedersen}\ \emph {et~al.}(1991)\citenamefont
  {Pedersen}, \citenamefont {Bj{\o}rnholm}, \citenamefont {Borggreen},
  \citenamefont {Hansen}, \citenamefont {Martin},\ and\ \citenamefont
  {Rasmussen}}]{PedersenNature1991}%
  \BibitemOpen
  \bibfield  {author} {\bibinfo {author} {\bibfnamefont {J.}~\bibnamefont
  {Pedersen}}, \bibinfo {author} {\bibfnamefont {S.}~\bibnamefont
  {Bj{\o}rnholm}}, \bibinfo {author} {\bibfnamefont {J.}~\bibnamefont
  {Borggreen}}, \bibinfo {author} {\bibfnamefont {K.}~\bibnamefont {Hansen}},
  \bibinfo {author} {\bibfnamefont {T.~P.}\ \bibnamefont {Martin}},\ and\
  \bibinfo {author} {\bibfnamefont {H.}~\bibnamefont {Rasmussen}},\ }\href@noop
  {} {\bibfield  {journal} {\bibinfo  {journal} {Nature}\ }\textbf {\bibinfo
  {volume} {353}},\ \bibinfo {pages} {733} (\bibinfo {year}
  {1991})}\BibitemShut {NoStop}%
\bibitem [{\citenamefont {Saunders}\ \emph {et~al.}(1985)\citenamefont
  {Saunders}, \citenamefont {Clemenger}, \citenamefont {de~Heer},\ and\
  \citenamefont {Knight}}]{SaundersPRB1985}%
  \BibitemOpen
  \bibfield  {author} {\bibinfo {author} {\bibfnamefont {W.}~\bibnamefont
  {Saunders}}, \bibinfo {author} {\bibfnamefont {K.}~\bibnamefont {Clemenger}},
  \bibinfo {author} {\bibfnamefont {W.}~\bibnamefont {de~Heer}},\ and\ \bibinfo
  {author} {\bibfnamefont {W.}~\bibnamefont {Knight}},\ }\href@noop {}
  {\bibfield  {journal} {\bibinfo  {journal} {Phys. Rev. B}\ }\textbf {\bibinfo
  {volume} {32}},\ \bibinfo {pages} {1366} (\bibinfo {year}
  {1985})}\BibitemShut {NoStop}%
\bibitem [{\citenamefont {Manninen}\ \emph {et~al.}(1994)\citenamefont
  {Manninen}, \citenamefont {Mansikka-aho}, \citenamefont {Nishioka},\ and\
  \citenamefont {Takahashi}}]{ManninenZPD1994}%
  \BibitemOpen
  \bibfield  {author} {\bibinfo {author} {\bibfnamefont {M.}~\bibnamefont
  {Manninen}}, \bibinfo {author} {\bibfnamefont {J.}~\bibnamefont
  {Mansikka-aho}}, \bibinfo {author} {\bibfnamefont {H.}~\bibnamefont
  {Nishioka}},\ and\ \bibinfo {author} {\bibfnamefont {Y.}~\bibnamefont
  {Takahashi}},\ }\href@noop {} {\bibfield  {journal} {\bibinfo  {journal} {Z.
  Phys. D}\ }\textbf {\bibinfo {volume} {31}},\ \bibinfo {pages} {259}
  (\bibinfo {year} {1994})}\BibitemShut {NoStop}%
\bibitem [{\citenamefont {Andersen}\ \emph {et~al.}(1996)\citenamefont
  {Andersen}, \citenamefont {Brink}, \citenamefont {Hvelplund}, \citenamefont
  {Larsson}, \citenamefont {Nielsen},\ and\ \citenamefont
  {Shen}}]{Andersen1996}%
  \BibitemOpen
  \bibfield  {author} {\bibinfo {author} {\bibfnamefont {J.~U.}\ \bibnamefont
  {Andersen}}, \bibinfo {author} {\bibfnamefont {C.}~\bibnamefont {Brink}},
  \bibinfo {author} {\bibfnamefont {P.}~\bibnamefont {Hvelplund}}, \bibinfo
  {author} {\bibfnamefont {M.~O.}\ \bibnamefont {Larsson}}, \bibinfo {author}
  {\bibfnamefont {B.~B.}\ \bibnamefont {Nielsen}},\ and\ \bibinfo {author}
  {\bibfnamefont {H.}~\bibnamefont {Shen}},\ }\href
  {https://doi.org/10.1103/PhysRevLett.77.3991} {\bibfield  {journal} {\bibinfo
   {journal} {Phys. Rev. Lett.}\ }\textbf {\bibinfo {volume} {77}},\ \bibinfo
  {pages} {3991} (\bibinfo {year} {1996})}\BibitemShut {NoStop}%
\bibitem [{\citenamefont {Andersen}\ \emph {et~al.}(2001)\citenamefont
  {Andersen}, \citenamefont {Gottrup}, \citenamefont {Hansen}, \citenamefont
  {Hvelplund},\ and\ \citenamefont {Larsson}}]{AndersenEPJD2001}%
  \BibitemOpen
  \bibfield  {author} {\bibinfo {author} {\bibfnamefont {J.}~\bibnamefont
  {Andersen}}, \bibinfo {author} {\bibfnamefont {C.}~\bibnamefont {Gottrup}},
  \bibinfo {author} {\bibfnamefont {K.}~\bibnamefont {Hansen}}, \bibinfo
  {author} {\bibfnamefont {P.}~\bibnamefont {Hvelplund}},\ and\ \bibinfo
  {author} {\bibfnamefont {M.~O.}\ \bibnamefont {Larsson}},\ }\href@noop {}
  {\bibfield  {journal} {\bibinfo  {journal} {Eur. Phys. J. D}\ }\textbf
  {\bibinfo {volume} {17}},\ \bibinfo {pages} {189} (\bibinfo {year}
  {2001})}\BibitemShut {NoStop}%
\bibitem [{\citenamefont {Toker}\ \emph {et~al.}(2007)\citenamefont {Toker},
  \citenamefont {Aviv}, \citenamefont {Eritt}, \citenamefont {Rappaport},
  \citenamefont {Heber}, \citenamefont {Schwalm},\ and\ \citenamefont
  {Zajfman}}]{toker07}%
  \BibitemOpen
  \bibfield  {author} {\bibinfo {author} {\bibfnamefont {Y.}~\bibnamefont
  {Toker}}, \bibinfo {author} {\bibfnamefont {O.}~\bibnamefont {Aviv}},
  \bibinfo {author} {\bibfnamefont {M.}~\bibnamefont {Eritt}}, \bibinfo
  {author} {\bibfnamefont {M.~L.}\ \bibnamefont {Rappaport}}, \bibinfo {author}
  {\bibfnamefont {O.}~\bibnamefont {Heber}}, \bibinfo {author} {\bibfnamefont
  {D.}~\bibnamefont {Schwalm}},\ and\ \bibinfo {author} {\bibfnamefont
  {D.}~\bibnamefont {Zajfman}},\ }\href
  {https://doi.org/10.1103/PhysRevA.76.053201} {\bibfield  {journal} {\bibinfo
  {journal} {Phys. Rev. A}\ }\textbf {\bibinfo {volume} {76}},\ \bibinfo
  {pages} {053201} (\bibinfo {year} {2007})}\BibitemShut {NoStop}%
\bibitem [{\citenamefont {Martin}\ \emph {et~al.}(2013)\citenamefont {Martin},
  \citenamefont {Bernard}, \citenamefont {{Br{\'e}dy}}, \citenamefont
  {Concina}, \citenamefont {Joblin}, \citenamefont {Ji}, \citenamefont
  {Ort{\'e}ga},\ and\ \citenamefont {Chen}}]{martin13}%
  \BibitemOpen
  \bibfield  {author} {\bibinfo {author} {\bibfnamefont {S.}~\bibnamefont
  {Martin}}, \bibinfo {author} {\bibfnamefont {J.}~\bibnamefont {Bernard}},
  \bibinfo {author} {\bibfnamefont {R.}~\bibnamefont {{Br{\'e}dy}}}, \bibinfo
  {author} {\bibfnamefont {B.}~\bibnamefont {Concina}}, \bibinfo {author}
  {\bibfnamefont {C.}~\bibnamefont {Joblin}}, \bibinfo {author} {\bibfnamefont
  {M.}~\bibnamefont {Ji}}, \bibinfo {author} {\bibfnamefont {C.}~\bibnamefont
  {Ort{\'e}ga}},\ and\ \bibinfo {author} {\bibfnamefont {L.}~\bibnamefont
  {Chen}},\ }\href {https://doi.org/10.1103/PhysRevLett.110.063003} {\bibfield
  {journal} {\bibinfo  {journal} {Phys. Rev. Lett.}\ }\textbf {\bibinfo
  {volume} {110}},\ \bibinfo {pages} {063003} (\bibinfo {year}
  {2013})}\BibitemShut {NoStop}%
\bibitem [{\citenamefont {Ebara}\ \emph {et~al.}(2016)\citenamefont {Ebara},
  \citenamefont {Furukawa}, \citenamefont {Matsumoto}, \citenamefont {Tanuma},
  \citenamefont {Azuma}, \citenamefont {Shiromaru},\ and\ \citenamefont
  {Hansen}}]{EbaraPRL2016}%
  \BibitemOpen
  \bibfield  {author} {\bibinfo {author} {\bibfnamefont {Y.}~\bibnamefont
  {Ebara}}, \bibinfo {author} {\bibfnamefont {T.}~\bibnamefont {Furukawa}},
  \bibinfo {author} {\bibfnamefont {J.}~\bibnamefont {Matsumoto}}, \bibinfo
  {author} {\bibfnamefont {H.}~\bibnamefont {Tanuma}}, \bibinfo {author}
  {\bibfnamefont {T.}~\bibnamefont {Azuma}}, \bibinfo {author} {\bibfnamefont
  {H.}~\bibnamefont {Shiromaru}},\ and\ \bibinfo {author} {\bibfnamefont
  {K.}~\bibnamefont {Hansen}},\ }\href@noop {} {\bibfield  {journal} {\bibinfo
  {journal} {Phys. Rev. Lett.}\ }\textbf {\bibinfo {volume} {117}},\ \bibinfo
  {pages} {133004} (\bibinfo {year} {2016})}\BibitemShut {NoStop}%
\bibitem [{\citenamefont {Ferrari}\ \emph {et~al.}(2019)\citenamefont
  {Ferrari}, \citenamefont {Janssens}, \citenamefont {Lievens},\ and\
  \citenamefont {Hansen}}]{FerrariIRPC2019}%
  \BibitemOpen
  \bibfield  {author} {\bibinfo {author} {\bibfnamefont {P.}~\bibnamefont
  {Ferrari}}, \bibinfo {author} {\bibfnamefont {E.}~\bibnamefont {Janssens}},
  \bibinfo {author} {\bibfnamefont {P.}~\bibnamefont {Lievens}},\ and\ \bibinfo
  {author} {\bibfnamefont {K.}~\bibnamefont {Hansen}},\ }\href@noop {}
  {\bibfield  {journal} {\bibinfo  {journal} {Int. Rev. Phys. Chem.}\ }\textbf
  {\bibinfo {volume} {38}},\ \bibinfo {pages} {405} (\bibinfo {year}
  {2019})}\BibitemShut {NoStop}%
\bibitem [{\citenamefont {Saito}\ \emph {et~al.}(2020)\citenamefont {Saito},
  \citenamefont {Kubota}, \citenamefont {Yamasa}, \citenamefont {Suzuki},
  \citenamefont {Majima},\ and\ \citenamefont {Tsuchida}}]{SaitoPRA2020}%
  \BibitemOpen
  \bibfield  {author} {\bibinfo {author} {\bibfnamefont {M.}~\bibnamefont
  {Saito}}, \bibinfo {author} {\bibfnamefont {H.}~\bibnamefont {Kubota}},
  \bibinfo {author} {\bibfnamefont {K.}~\bibnamefont {Yamasa}}, \bibinfo
  {author} {\bibfnamefont {K.}~\bibnamefont {Suzuki}}, \bibinfo {author}
  {\bibfnamefont {T.}~\bibnamefont {Majima}},\ and\ \bibinfo {author}
  {\bibfnamefont {H.}~\bibnamefont {Tsuchida}},\ }\href
  {https://doi.org/10.1103/PhysRevA.102.012820} {\bibfield  {journal} {\bibinfo
   {journal} {Phys. Rev. A}\ }\textbf {\bibinfo {volume} {102}},\ \bibinfo
  {pages} {012820} (\bibinfo {year} {2020})}\BibitemShut {NoStop}%
\bibitem [{\citenamefont {Kubo}(1962)}]{Kubo}%
  \BibitemOpen
  \bibfield  {author} {\bibinfo {author} {\bibfnamefont {R.}~\bibnamefont
  {Kubo}},\ }\href@noop {} {\bibfield  {journal} {\bibinfo  {journal} {J. Phys.
  Soc. Japan}\ }\textbf {\bibinfo {volume} {17}},\ \bibinfo {pages} {975}
  (\bibinfo {year} {1962})}\BibitemShut {NoStop}%
\bibitem [{\citenamefont {Denton}\ \emph {et~al.}(1973)\citenamefont {Denton},
  \citenamefont {M{\"u}hlschlegel},\ and\ \citenamefont
  {Scalapino}}]{DentonPRB1973}%
  \BibitemOpen
  \bibfield  {author} {\bibinfo {author} {\bibfnamefont {R.}~\bibnamefont
  {Denton}}, \bibinfo {author} {\bibfnamefont {B.}~\bibnamefont
  {M{\"u}hlschlegel}},\ and\ \bibinfo {author} {\bibfnamefont {D.~J.}\
  \bibnamefont {Scalapino}},\ }\href@noop {} {\bibfield  {journal} {\bibinfo
  {journal} {Phys. Rev. B}\ }\textbf {\bibinfo {volume} {7}},\ \bibinfo {pages}
  {3589} (\bibinfo {year} {1973})}\BibitemShut {NoStop}%
\bibitem [{\citenamefont {Brack}\ \emph {et~al.}(1991)\citenamefont {Brack},
  \citenamefont {Genzken},\ and\ \citenamefont {Hansen}}]{BrackZPD1991}%
  \BibitemOpen
  \bibfield  {author} {\bibinfo {author} {\bibfnamefont {M.}~\bibnamefont
  {Brack}}, \bibinfo {author} {\bibfnamefont {O.}~\bibnamefont {Genzken}},\
  and\ \bibinfo {author} {\bibfnamefont {K.}~\bibnamefont {Hansen}},\
  }\href@noop {} {\bibfield  {journal} {\bibinfo  {journal} {Z. Phys. D}\
  }\textbf {\bibinfo {volume} {21}},\ \bibinfo {pages} {65} (\bibinfo {year}
  {1991})}\BibitemShut {NoStop}%
\bibitem [{\citenamefont {Hansen}(2020)}]{HansenCP2020}%
  \BibitemOpen
  \bibfield  {author} {\bibinfo {author} {\bibfnamefont {K.}~\bibnamefont
  {Hansen}},\ }\href@noop {} {\bibfield  {journal} {\bibinfo  {journal} {Chem.
  Phys.}\ }\textbf {\bibinfo {volume} {530}},\ \bibinfo {pages} {110637(1}
  (\bibinfo {year} {2020})}\BibitemShut {NoStop}%
\bibitem [{\citenamefont {Frederick~C.Williams}(1969)}]{WilliamsNPA1969}%
  \BibitemOpen
  \bibfield  {author} {\bibinfo {author} {\bibfnamefont {J.}~\bibnamefont
  {Frederick~C.Williams}},\ }\href@noop {} {\bibfield  {journal} {\bibinfo
  {journal} {Nuclear Physics A}\ }\textbf {\bibinfo {volume} {133}},\ \bibinfo
  {pages} {33} (\bibinfo {year} {1969})}\BibitemShut {NoStop}%
\bibitem [{\citenamefont {Borrmann}\ and\ \citenamefont
  {Franke}(1993)}]{BorrmannJCP1993}%
  \BibitemOpen
  \bibfield  {author} {\bibinfo {author} {\bibfnamefont {P.}~\bibnamefont
  {Borrmann}}\ and\ \bibinfo {author} {\bibfnamefont {G.}~\bibnamefont
  {Franke}},\ }\href@noop {} {\bibfield  {journal} {\bibinfo  {journal} {J.
  Chem. Phys.}\ }\textbf {\bibinfo {volume} {98}},\ \bibinfo {pages} {2484}
  (\bibinfo {year} {1993})}\BibitemShut {NoStop}%
\bibitem [{\citenamefont {Sch\"{o}nhammer}(2000)}]{SchonhammerAJP2000}%
  \BibitemOpen
  \bibfield  {author} {\bibinfo {author} {\bibfnamefont {K.}~\bibnamefont
  {Sch\"{o}nhammer}},\ }\href@noop {} {\bibfield  {journal} {\bibinfo
  {journal} {Am. J. Phys.}\ }\textbf {\bibinfo {volume} {68}},\ \bibinfo
  {pages} {1032} (\bibinfo {year} {2000})}\BibitemShut {NoStop}%
\end{thebibliography}
\end{document}